\documentclass[aps,amsmath,amssymb,nofootinbib,preprint numbers,twocolumn]{revtex4} 
\usepackage{graphicx} 
\usepackage{amsmath}
\usepackage{amsfonts,amsbsy}
\usepackage{amssymb}

\allowdisplaybreaks

\usepackage{xcolor}
\definecolor{lcolor}{rgb}{0.5,0,0}
\definecolor{citcolor}{rgb}{0,0.3,0.0}
\usepackage[breaklinks,colorlinks,urlcolor=blue,citecolor=citcolor,linkcolor=lcolor]{hyperref}

\usepackage{mciteplus}

\def\gsim{ \,\, \vcenter{\hbox{$\buildrel{\displaystyle >}\over\sim$}}
 \,\,}
\def\lsim{ \,\, \vcenter{\hbox{$\buildrel{\displaystyle <}\over\sim$}}
 \,\,}
\def\be{\begin{equation}}
\def\ee{\end{equation}}
\def\bea{\begin{eqnarray}}
\def\eea{\end{eqnarray}}

\newcommand{\dd}{{\rm d}}
\newcommand{\nn}{\nonumber}

\usepackage{subfig}

\begin{document}

\title{Erratum: Stronger $C$-odd color charge correlations in the proton at higher energy}

\author{Adrian Dumitru}
\affiliation{Department of Natural Sciences, Baruch College, CUNY 17 Lexington Avenue, New York, NY 10010, USA}
\affiliation{The Graduate School and University Center, The City University of New York, 365 Fifth Avenue, New York, NY 10016, USA}
\author{Heikki M\"{a}ntysaari}
\email{heikki.mantysaari@jyu.fi}
\affiliation{
Department of Physics, University of Jyväskylä,  P.O. Box 35, 40014 University of Jyväskylä, Finland
}
\affiliation{
Helsinki Institute of Physics, P.O. Box 64, 00014 University of Helsinki, Finland
}
\author{Risto Paatelainen}
\affiliation{
Helsinki Institute of Physics, P.O. Box 64, 00014 University of Helsinki, Finland
}

\maketitle

In the calculation of the $\left\langle \rho^a(\vec q_1)\rho^b(\vec q_2)\rho^c(\vec q_3)\right\rangle_{C=-1}$ correlator in Ref.~\cite{Dumitru:2021tqp} the symmetry factors for the diagrams of the ``Fig. 7 type'' included an incorrect factor $2$, see Erratum of Ref.~\cite{Dumitru:2021tqp}.

In this work we have evaluated the expressions in Ref.~\cite{Dumitru:2021tqp} numerically in order to calculate the C-odd odderon amplitude in coordinate space. Once the fixed symmetry factors are taken into account, the numerical results presented in this letter change. Furthermore, we also realized that an overall factor $(-1)$ was missing from all our results.

When the fixed symmetry factors are taken into account, the original conclusions change to some extent. We originally reported finding a stronger Odderon amplitude with decreasing Bjorken-$x$. This remains true for small dipoles, compared to the impact parameter. However, at larger dipole size $r$ the $C$-odd amplitude may either increase or decrease, depending on the impact parameter.

\begin{figure*}[tb]  
    \subfloat[Large dipole, small impact parameter]{%
        \includegraphics[width=\columnwidth]{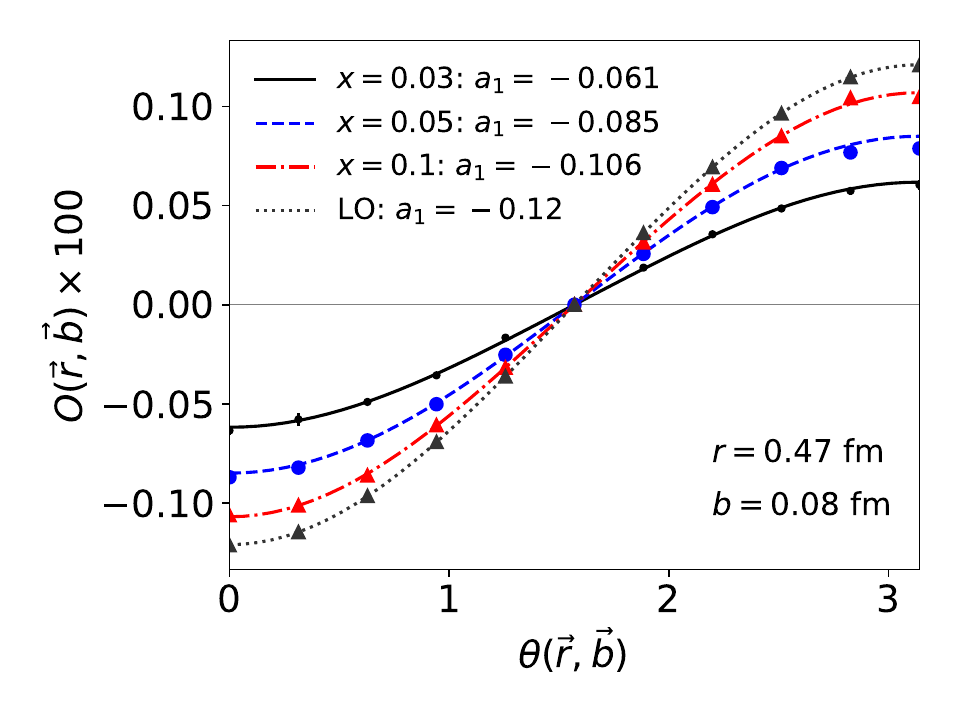}%
       \label{fig:a1_bdep}
    }
        \hfill
        \subfloat[Small dipole, large impact parameter]{
         \includegraphics[width=\columnwidth]{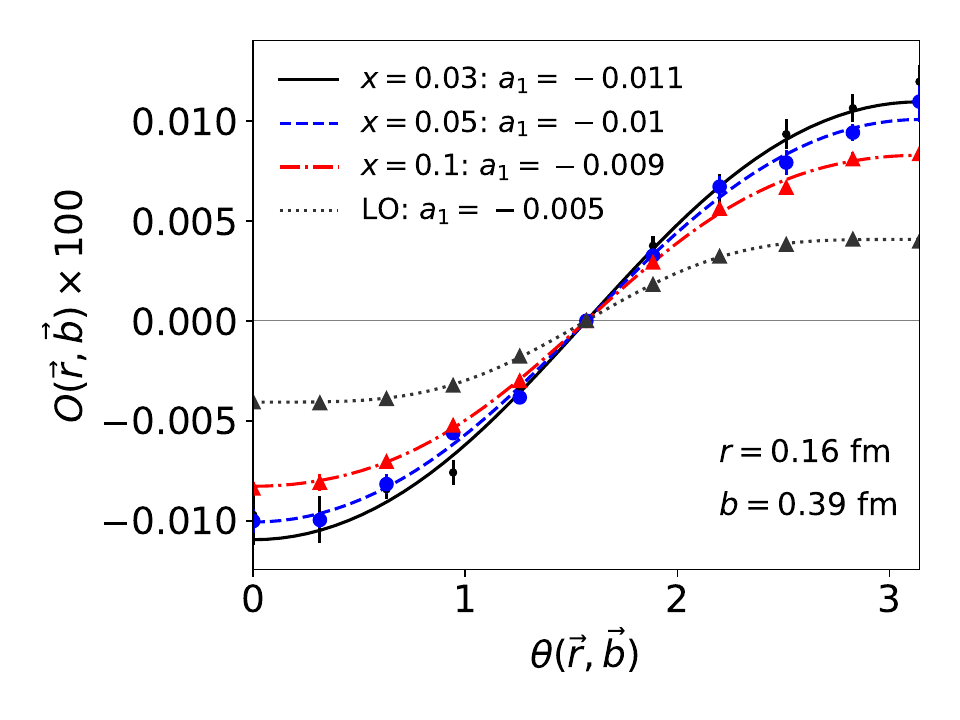}
          \label{fig:a1_rdep}
        }
    
    \caption{Angular dependence of $O(\vec r, \vec b)$ at various $x$. The error bars show the estimated uncertainty of the numerical Monte Carlo integration.}
    \label{fig:angledep}
    
\end{figure*}

We now present an updated numerical analysis of the odderon coordinate space amplitude. First, the angular dependence of the Odderon (defined in Eq. (3) of the original manusript) is shown in Fig.~\ref{fig:angledep}. The Odderon is parametrized as 
\begin{equation}
    O(\vec r,\vec b) = a_1(r,b) \cos \theta + a_3(r,b)\cos 3\theta
\end{equation}
and the dominant coefficient $a_1$ is shown in the following Figures. Depending on the dipole size and on the impact parameter, the NLO correction can be quite large already at $x=0.1$, and it in general exhibits a relatively moderate $x$ dependence.


The impact parameter dependence of the coefficient $a_1$ quantifying the magnitude of the Odderon is shown in Fig.~\ref{fig:a1_bdep}. As already mentioned above, the NLO correction now reduces the amplitude at small $b\ll r$,  while a stronger Odderon is obtained with decreasing $x$ at larger impact parameters.

\begin{figure*}[tb]  
    \subfloat[Small dipole]{%
        \includegraphics[width=\columnwidth]{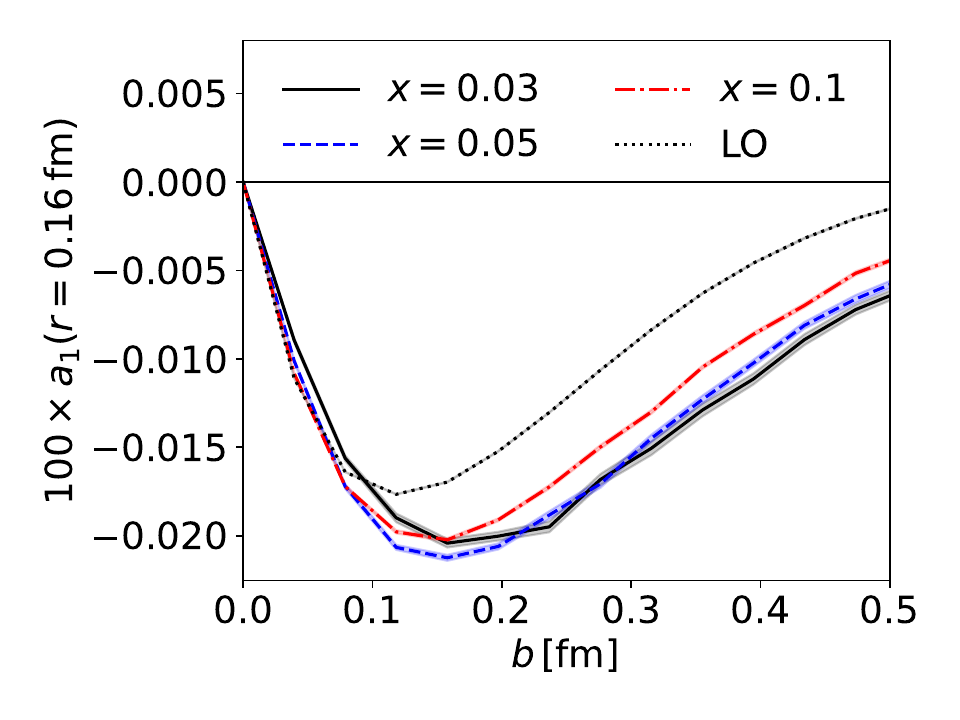}%
       
    }
        \hfill
        \subfloat[Large dipole]{
         \includegraphics[width=\columnwidth]{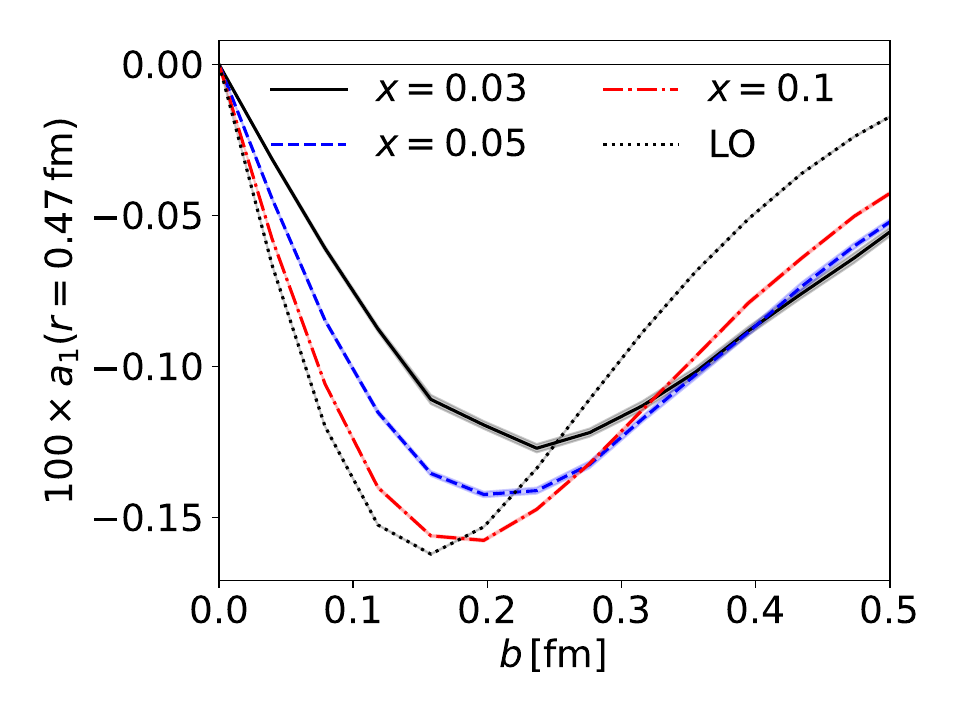}
          
        }
    
    \caption{Dependence on the impact parameter.}
    \label{fig:a1_bdep}
\end{figure*}

Similarly the dependence on the dipole size $r$ is shown in Fig.~\ref{fig:a1_rdep}, now at small and large $b$. Again a decreasing (in magnitude) Odderon with decreasing $x$ is obtained at small impact parameter, consistent with expectations based on perturbative small-$x$ evolution of the (hard) odderon~\cite{Lappi:2016gqe}. On the other hand, at larger $b$ the behavior is opposite.

In the original publication we reported that $a_1$ levels off at large $r \simeq 0.7$~fm. Such behavior is still visible
in Fig.~\ref{fig:a1_rdep}(left) for small $b$ (although we now restrict to $r\le 0.6$~fm where the numerical evaluation is
most reliable). At greater impact parameter $b$, however, see Fig.~\ref{fig:a1_rdep}(right), no levelling off of the
Odderon amplitude is seen up to the largest dipoles considered here. The behavior of $a_1(r)$ at different impact
parameters is also visible from Fig.~\ref{fig:grids}.

\begin{figure*}[tb]  
    \subfloat[Small impact parameter]{%
        \includegraphics[width=\columnwidth]{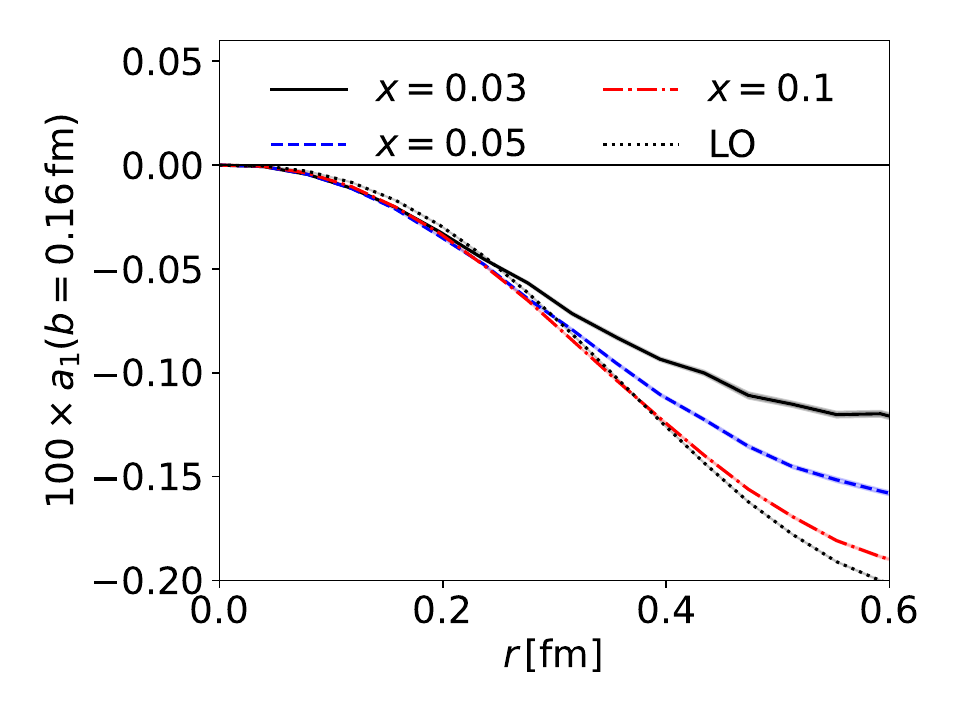}%
       
    }
        \hfill
        \subfloat[Large impact parameter]{
         \includegraphics[width=\columnwidth]{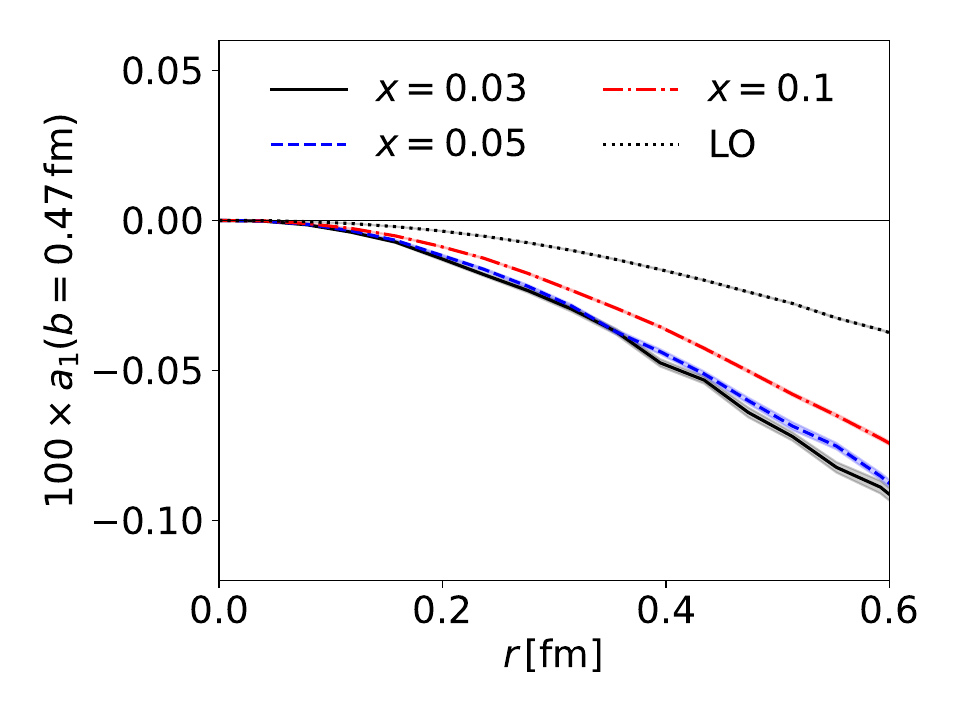}
          
        }
    
    \caption{Dependence on the dipole size.}
    \label{fig:a1_rdep}
\end{figure*}

We also provide updated two-dimensional grids for the coefficients $a_1$ and $a_3$ at different $r$. The grids for $a_1$ are illustrated in Fig.~\ref{fig:grids} and numerical values can be found in the supplementary material. 

\begin{figure*}[tb]  
    \subfloat[NLO $x=0.1$]{%
        \includegraphics[width=0.5\textwidth]{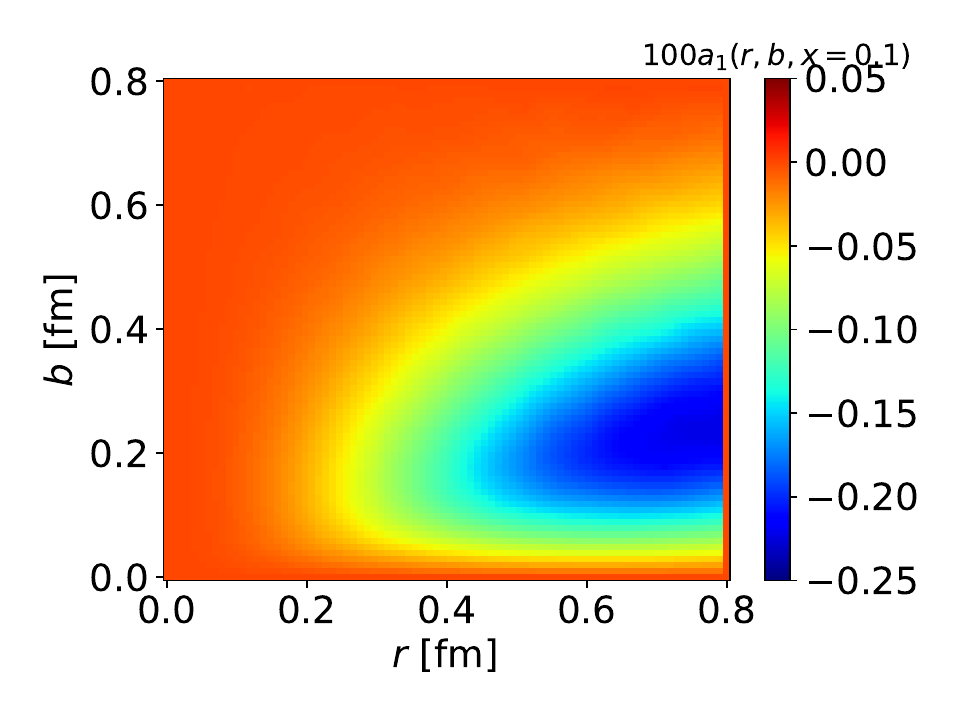}%
        \label{fig:}%
    }
    \hfill
    \subfloat[NLO $x=0.03$]{%
        \includegraphics[width=0.5\textwidth]{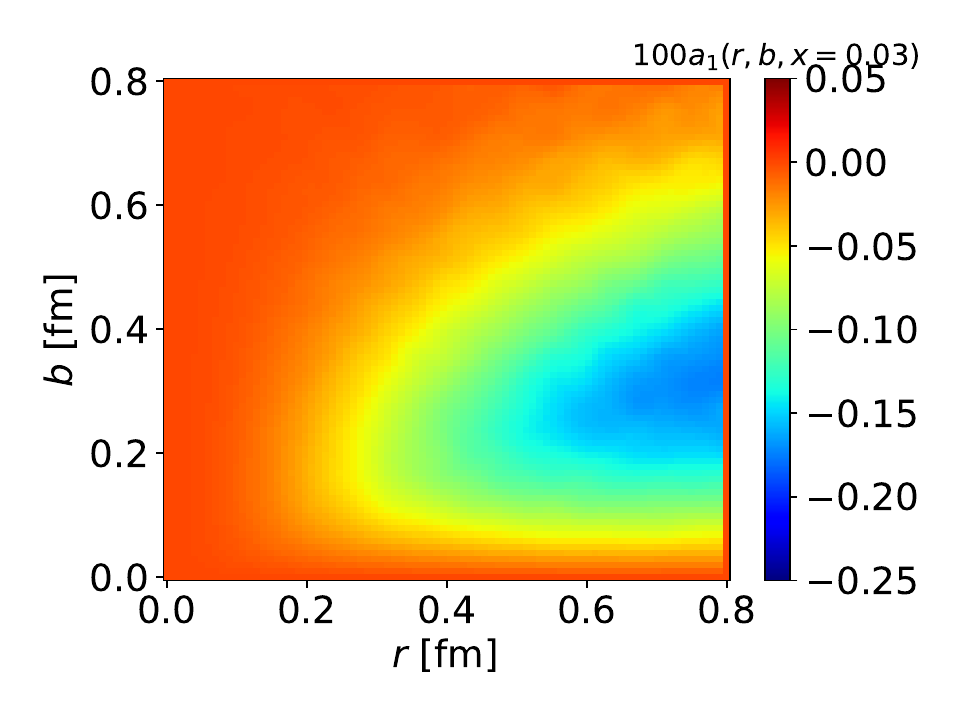}%
        \label{fig:}%
    }
    \caption{Odderon modulation coefficient $a_1$ as a function of dipole size $r$ and impact parameter $b$ at NLO.}
    \label{fig:grids}
\end{figure*}


\clearpage









\title{Stronger $C$-odd color charge correlations in the proton at higher
energy}

\begin{widetext}
   { \Large \bf Stronger $C$-odd color charge correlations in the proton at higher
energy }
\mbox{}\\

{\centering 
Adrian Dumitru\\
\emph{\small Department of Natural Sciences, Baruch College,
CUNY 17 Lexington Avenue, New York, NY 10010, USA and\\
The Graduate School and University Center, The City University of New York, 365 Fifth Avenue, New York, NY 10016, USA}\\
\mbox{}\\
Heikki Mäntysaari\\
\emph{\small Department of Physics, University of Jyväskylä, P.O. Box 35, 40014 University of Jyväskylä, Finland and\\
Helsinki Institute of Physics, P.O. Box 64, 00014 University of Helsinki, Finland}\\
\mbox{}\\
Risto Paatelainen\\
\emph{\small Helsinki Institute of Physics, P.O. Box 64, 00014 University of Helsinki, Finland}

}

\subsection*{Abstract}
The non-forward eikonal scattering matrix for dipole-proton scattering at
high energy obtains an imaginary part due to a $C$-odd three gluon
exchange.  We present numerical estimates for the perturbative
Odderon amplitude as a function of dipole size, impact parameter,
their relative azimuthal angle, and light-cone momentum cutoff
$x$. The proton is approximated as $\psi_\mathrm{qqq}|qqq\rangle
+ \psi_\mathrm{qqqg}|qqqg\rangle$, where $\psi_\mathrm{qqq}$ is a
non-perturbative three quark model wave function while the gluon
emission is computed in light-cone perturbation theory.  We find
that the Odderon amplitude 
increases as $x$ decreases from 0.1 to
0.01.  At yet lower $x$, the reversal of this energy dependence
would reflect the onset of universal small-$x$ renormalization group evolution.

\end{widetext}

\preprint{HIP-2022-24/TH}

\author{Adrian Dumitru}
\email{adrian.dumitru@baruch.cuny.edu}
\affiliation{Department of Natural Sciences, Baruch College, CUNY,
17 Lexington Avenue, New York, NY 10010, USA}
\affiliation{The Graduate School and University Center, The City University
  of New York, 365 Fifth Avenue, New York, NY 10016, USA}

\author{Heikki Mäntysaari}
\email{heikki.mantysaari@jyu.fi}
\affiliation{
Department of Physics, University of Jyväskylä,  P.O. Box 35, 40014 University of Jyväskylä, Finland
}
\affiliation{
Helsinki Institute of Physics, P.O. Box 64, 00014 University of Helsinki, Finland
}

\author{Risto Paatelainen} \email{risto.sakari.paatelainen@cern.ch}
\affiliation{Helsinki Institute of Physics and Department of Physics, FI-00014 University of Helsinki, Finland}



\newcommand{\lqcd}{\Lambda_{\mathrm{QCD}}}
\newcommand{\as}{\alpha_{\mathrm{s}}}

\newcommand{\fig}{Fig.~}
\newcommand{\figs}{Figs.~}
\newcommand{\eq}{Eq.~}
\newcommand{\se}{Sec.~}
\newcommand{\eqs}{Eqs.~}

\newcommand{\llsim}{{\underset{\sim}{\ll}}}
\newcommand{\ggsim}{{\underset{\sim}{\gg}}}

\section{Introduction}

The $S$-matrix for high-energy eikonal scattering of a quark -
antiquark dipole off the proton
is~\cite{Nikolaev:1990ja,Mueller:1994gb,Mueller:2001fv, Kovchegov:2012mbw}
\be \label{eq:S_dipole_b}
   {\cal S} (\vec x,\vec y) =
   \frac{1}{N_c}\,\left \langle {\rm tr} \,U\left(\vec x\right)\,
     U^\dagger\left(\vec y\right)\right\rangle \, .
\ee
Below we shall also use the impact parameter $\vec b=(\vec x + \vec
y)/2$ and dipole (transverse) vectors $\vec r = \vec y - \vec x$ where
$\vec r$ points from the anti-quark to the quark.  The
$\langle\cdots\rangle$ brackets denote the matrix element between the
incoming proton state $|P^+,\vec P=0\rangle$ and the outgoing state
$\langle P^+,\vec K|$, where $\vec K$ denotes the proton transverse momentum.  Our sign convention for the coupling in the
covariant derivative, $D_\mu = \partial_\mu + ig A_\mu^a t^a$, follows
Ref.~\cite{Brodsky:1997de}. Hence, the path ordered exponential of the
field in covariant gauge (Wilson line) which represents the eikonal
scattering of the quark is
\be \label{eq:WilsonLines}
U(\vec x) = {\cal P} e^{-ig \int \mathrm{d}x^- A^{+a}(x^-,\vec x)\, t^a} ~.
\ee
Our convention for the Wilson line and for the dipole $S$-matrix
agrees with Ref.~\cite{Kovchegov:2013cva}. Others such as
Ref.~\cite{Yao:2018vcg} define ${\cal S} (\vec x,\vec y)$ with $U
\leftrightarrow U^\dagger$; however, they also take $\vec r = \vec x -
\vec y$, so in all, the sign for the imaginary part of the $S$-matrix
is the same.

Indeed, our focus here is on the imaginary part $O(\vec r,\vec b)$ of
the $S$-matrix, the so-called ``$b$-dependent Odderon'',
which starts out in perturbation theory as $C$-odd
three gluon exchange.  This amplitude is odd under $C$-conjugation,
i.e.  exchange of quark and anti-quark.
The relation of various Odderon amplitudes to Generalized
Transverse Momentum Dependent parton distributions (GTMDs)
has been elucidated in refs.~\cite{Zhou:2013gsa,Boer:2015pni,Yao:2018vcg,Boussarie:2019vmk,Hagiwara:2020mqb,Boer:2022njw}.

The $C$-odd three gluon exchange couples to cubic color charge
fluctuations in the proton~\cite{Dumitru:2018vpr},
\begin{widetext}
\begin{multline}
    \mathrm{Im}\, S (\vec r, \vec b) = O(\vec r, \vec b) =
- \frac{5}{18}\, g^6~\frac{1}{2}\, \frac{1}{3}
\int\limits_{q_1, q_2, q_3>q_\text{min}}
\frac{1}{q_1^2}\frac{1}{q_2^2}\frac{1}{q_3^2}\,
\sin(\vec b \cdot \vec K)\,
G_3^-(\vec q_1,\vec q_2,\vec q_3)\\
 \left[\sum_{i=1,2,3}\left(
\sin\left(\vec r\cdot \vec q_i + \frac{1}{2} \vec r\cdot \vec
    K\right)
  - \sin\left(\vec r\cdot \vec q_i^{\, \prime} + \frac{1}{2} \vec r\cdot \vec
    K'\right)    \right)
- \sin\left(\frac{{1}}{2}\vec r\cdot
  \vec K\right)
    +\sin\left(\frac{{1}}{2}\vec r\cdot
  \vec K'\right)\right]~,
\label{eq:Odderon_G3-}
\end{multline}
\end{widetext}
We have written $O(\vec r, \vec b)$ in a form which is more suitable for numerical integration, in particular the amplitude vanishes already at the integrand level when $\vec r \perp \vec b$ and different momenta $\vec q_i$ appear in a symmetric form.
The sign of $O(\vec r, \vec b)$ differs
from Ref.~\cite{Dumitru:2018vpr} because here we employ the more
common convention $\vec r = \vec y - \vec x$ rather than $\vec r =
\vec x - \vec y$.
Here the parameter $g = \sqrt{4\pi\as}$ is the strong coupling constant, $\vec K = - (\vec q_1 + \vec q_2 + \vec
q_3)$ is the transverse momentum transfer given $\vec P=0$ for the
incoming proton, and $\int_q$ is shorthand for $\int
\dd^2q/(2\pi)^2$. Also, the transverse momentum vectors $\vec q_i^{\,
  \prime}$ correspond to sign-flipped components along $\vec b$.  We
have also introduced a low momentum cutoff $q_\mathrm{min}$ for
numerical stability; no significant dependence on this cutoff was
observed when $q_\mathrm{min} < 0.1\,\mathrm{GeV}$, except in regions where
$O(\vec r, \vec b)$ has a very small magnitude. The actual numerical results shown in this paper are obtained using $q_\mathrm{min}=0.03\,\mathrm{GeV}$.

We denote the $C$-odd part of the light-cone gauge correlator of three
color charge operators as
\be
\left\langle
\rho^a(\vec q_1) \, \rho^b(\vec q_2)\, \rho^c(\vec q_3)\right\rangle_{C=-} \equiv
\frac{1}{4} d^{abc}\, g^3\, G_3^-(\vec q_1,\vec q_2,\vec q_3)~.
\ee
Here, $\rho^a$ corresponds to the plus component of the color current, integrated
over $x^-$. In terms of creation and annihilation operators for quarks and gluons it reads~\cite{Dumitru:2020gla}
\bea
\rho^a(\vec k) &=& g \sum_{i,j,\sigma} (t^a)_{ij}
\int\frac{\dd x_q \dd^2q}{16\pi^3\, x_q}
 b^\dagger_{i\sigma}(x_q,\vec q)\, b_{j\sigma}(x_q,\vec k+\vec q) \nn\\
 & & \hspace{-1cm} + g  \sum_{\lambda b c} (T^a)_{bc}
\int \frac{\dd x_g \dd^2q}{16\pi^3\, x_g}
a^\dagger_{b\lambda}(x_g,\vec q) \, a_{c\lambda}(x_g,\vec k+\vec q) \, .
\eea

Ref.~\cite{Dumitru:2020fdh} evaluated $G_3^-(\vec q_1,\vec q_2,\vec q_3)$ for a
non-perturbative three quark light-cone constituent quark
model~\cite{Schlumpf:1992vq,Brodsky:1994fz}. This model provides
realistic one-particle longitudinal and transverse momentum
distributions, and also encodes momentum correlations. We refer to 
this three-quark light-cone
wave function as the leading-order (LO) approximation.

The diagrams corresponding to
corrections to the impact factor due to the perturbative emission of a gluon have been
computed in Ref.~\cite{Dumitru:2021tqp}; they are too numerous to be
listed again here. This will be referred to as the next-to-leading order (NLO) approximation.
The purpose of this paper is to present numerical
results for $O(\vec r, \vec b)$ from this approach, which together
with analogous results for the real part of $S(\vec r, \vec
b)$~\cite{Dumitru:2020gla,Dumitru:2021tvw} provide a complete set of
initial conditions for small-$x$ evolution of the dipole $S$-matrix.
The questions we address here are about the overall magnitude of the
three gluon exchange amplitude, and its dependence on $r=|\vec r|$, $b=|\vec b|$, their
relative angle $\theta$, and on the cutoff $x$ on the parton light-cone
momentum which appears in $G_3^-$.

The non-vanishing imaginary part of the $S$-matrix
can be probed, for example, via charge asymmetries in diffractive
electroproduction of a $\pi^+\, \pi^-$
pair~\cite{Hagler:2002nh,Hagler:2002nf}, exclusive production
of a pseudo-scalar meson~\cite{Czyzewski:1996bv, Engel:1997cga,
  Kilian:1997ew, Rueter:1998gj, Dumitru:2019qec} in deeply-inelastic
scattering (DIS) or ultra-peripheral proton-nucleus collisions, lepton-meson azimuthal angle correlations in exclusive processes~\cite{Mantysaari:2020lhf}
as well as in exclusive production of a vector meson
in $p+p$ scattering~\cite{Bzdak:2007cz} via ``pomeron-odderon fusion''.

Finally, it is also our goal to provide numerical estimates for
initial conditions for small-$x$ QCD evolution of the (hard) Odderon
$O(\vec r,\vec b)$~\cite{Kovchegov:2003dm, Hatta:2005as,
  Lappi:2016gqe}. Their crude knowledge, see
e.g.\ Refs.~\cite{Lappi:2016gqe,Yao:2018vcg}, is a key
limitation for quantitative predictions of the observables mentioned
above in the energy regime of the Electron-Ion Collider
(EIC)~\cite{Accardi:2012qut,Aschenauer:2017jsk,AbdulKhalek:2021gbh}.

\section{Results}

The results presented here apply when the $C$-odd exchange can be
described by the exchange of three gluons, i.e.\ in the perturbative
regime. This should be the case when the scattered dipole is small
and/or when the momentum transfer (conjugate to the impact parameter)
is large. Furthermore, since we only consider the $|qqq\rangle$ and
$|qqqg\rangle$ Fock states of the proton, we restrict to $x \gsim
0.01$.  The results shown below have been obtained with
$\alpha_s= 0.2$; note that aside from the overall
$\alpha_s^3$ prefactor in Eq.~\eqref{eq:Odderon_G3-}, the NLO
contribution to $G_3^-$, too, depends on the coupling, see
Ref.~\cite{Dumitru:2021tqp}. Note also that the coupling does not run at this order as the perturbative one gluon emission corrections are $\mathcal{O}(\alpha_s)$. 

The non-perturbative three-quark wave function for the proton used in the numerical analysis is the ``harmonic oscillator'' wave function of Ref.~\cite{Brodsky:1994fz}.  It has been used previously in Refs.~\cite{Dumitru:2021tqp,Dumitru:2021tvw} for estimates of the real part of the $S$-matrix. The parameters of the wave function are constrained by the 
proton radius, the anomalous magnetic moment and the axial coupling of the proton and the neutron. 
Given these constraints,  color charge correlators are not very sensitive to the particular model of the three-quark wave function~\cite{Dumitru:2021tvw}. Also, following Ref.~\cite{Dumitru:2021tvw},
here we evaluate all diagrams for the three gluon exchange with a collinear regulator of $m_\mathrm{col}=0.2\,\mathrm{GeV}$; this is consistent with the typical quark transverse momentum
in the wave function of Refs.~\cite{Schlumpf:1992vq,Brodsky:1994fz}.

\begin{figure}
    \centering
    \includegraphics[width=\columnwidth]{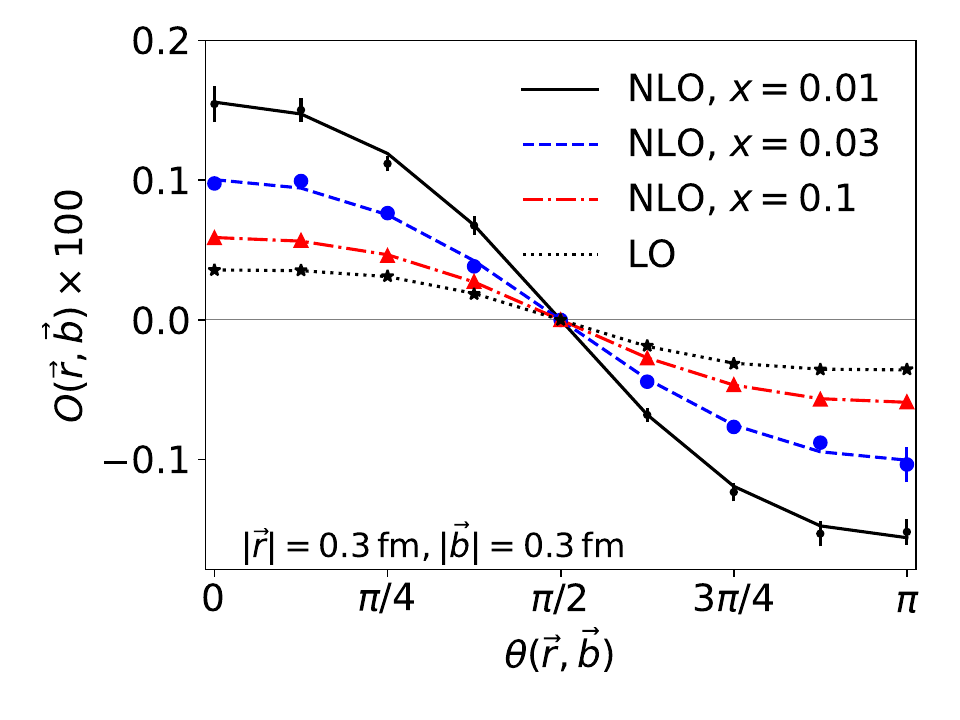}
    \caption{
    Angular dependence of $O(\vec r, \vec b)$ at various $x$ and $r=b=0.3$~fm, which is predominantly
    $\sim \hat r \cdot \hat b$. The coefficients (scaled by 100) are $a_1=-0.16, a_3=0.0063$ at $x=0.01$, $a_1=-0.10, a_3=0.0030$ at $x=0.03$ and $a_1=-0.063, a_3=0.0035$ at $x=0.1$. For comparison: at leading order the fitted coefficients are $a_1=-0.040$ and $a_3=0.0040$. The error bars show the estimated uncertainty of the numerical Monte Carlo integration.}
    \label{fig:angledep1}
\end{figure}

At the level of accuracy that we achieved in evaluating
Eq.~\eqref{eq:Odderon_G3-} we found that the angular dependence of
the odderon amplitude is well approximated by
\be
O(\vec r, \vec b) = a_1(r,b)\, \cos\theta +
a_3(r,b)\, \cos 3\theta ~,
\label{eq:expansion}
\ee
where $\theta$ is the azimuthal angle made by $\vec b$ and $\vec r$.
We typically find that the magnitude of $a_3$ is much smaller than that of $a_1$
except in the vicinity of a sign change of $a_1(r,b)$
where $O(\vec r, \vec b)$ is small.  The angular dependence of the odderon amplitude at $r=b=0.3\,\mathrm{fm}$ is shown in Fig.~\ref{fig:angledep1}. The amplitude obtained from the leading order calculation, where the dependence on the parton momentum fraction cutoff $x$ is negligible, is compared to the result of the NLO computation at $x=0.1,x=0.03$ and $x=0.01$.

These results show the correction due to the
perturbative gluon for different values of $x$. At $x=0.1$ this
correction is moderate, visible mostly for (anti-)parallel
$\vec r$ and $\vec b$,
as the phase space for gluon emission is
restricted. Note that the Odderon amplitude vanishes exactly when $\theta=0$ as can be seen from Eq.~\eqref{eq:Odderon_G3-}. For smaller $x$, although the qualitative angular
dependence remains the same, we observe a considerable
{\em increase} of the Odderon amplitude $|O(\vec r, \vec b)|$.

\begin{figure}
    \centering
    \includegraphics[width=\columnwidth]{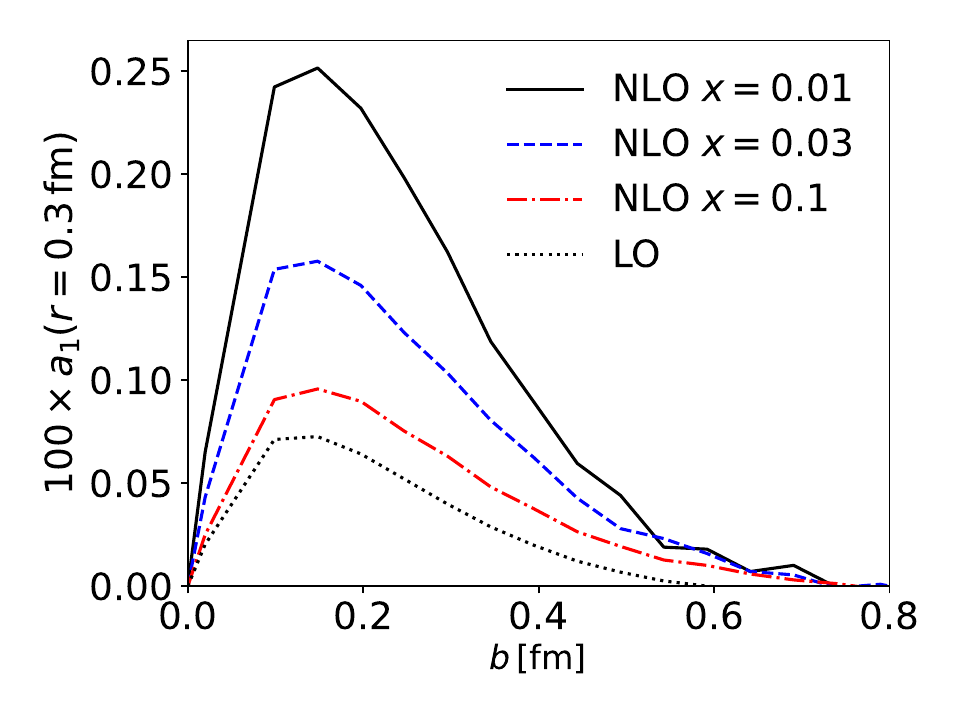}
    \caption{Impact parameter dependence of the odderon amplitude modulation coefficient $a_1$ defined in Eq.~\eqref{eq:expansion}. }
    \label{fig:bdep}
\end{figure}

\begin{figure}
    \centering
    \includegraphics[width=\columnwidth]{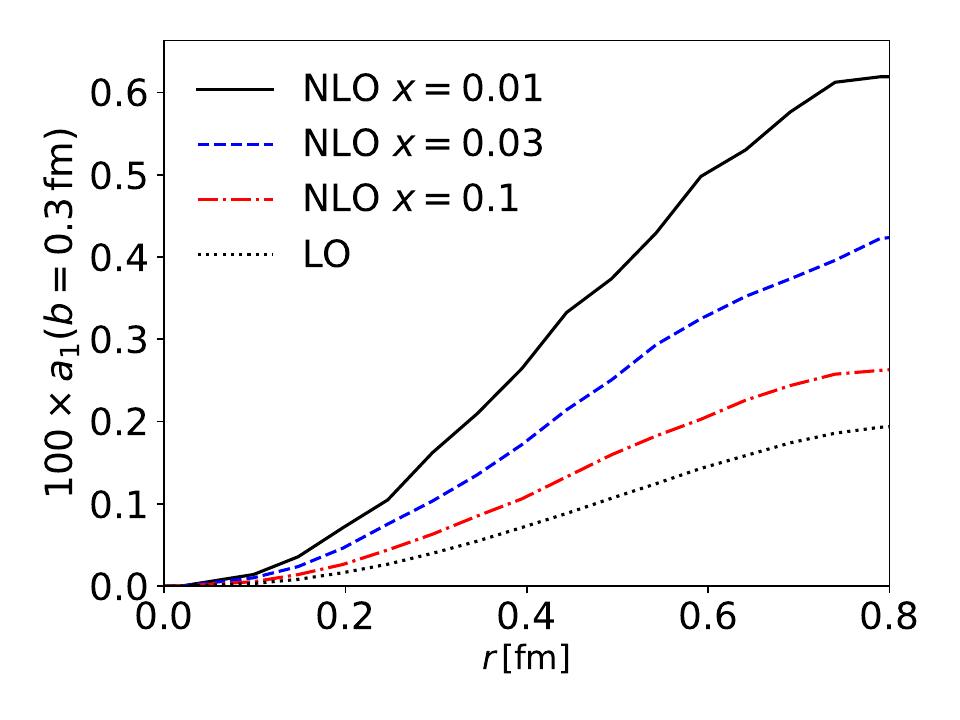}
    \caption{Dipole size dependence of $a_1$ at $b=0.3$~fm, and various $x$.}
    \label{fig:rdep}
\end{figure}

To further demonstrate the role of the NLO corrections on the Odderon amplitude, we show in Figs.~\ref{fig:bdep} and~\ref{fig:rdep} the dominant $a_1$ coefficient as a function of impact parameter (Fig.~\ref{fig:bdep}) and dipole size (Fig.~\ref{fig:rdep}). The next-to-leading order amplitudes computed at different longitudinal momentum fraction cutoffs $x$ are compared with the leading order result.
The Odderon amplitude is parity odd and so it vanishes at $b=0$. It increases with impact parameter and peaks at $b$ slightly less
than 0.2~fm, for a dipole size $r=0.3$~fm, followed by a smooth fall-off towards large $b$. The peak at $b\lsim{0.2}$~fm is seen
at much smaller scales than the transverse size $\sqrt{\langle b^2\rangle} \simeq 0.6$~fm associated with the real part
of the $S$-matrix extracted from fits to HERA data on exclusive $J/\Psi$ production in DIS~\cite{Kowalski:2006hc}. The peak position depends weakly on $r$ but remains at $b\lesssim 0.3$~fm for all dipole sizes $r\lesssim 0.8$~fm considered here.
Again we notice that the qualitative
shape of $a_1(b)$ is preserved by the NLO correction. However, while this correction is moderate at $x=0.1$, it
increases strongly with decreasing $x$.

Fig.~\ref{fig:rdep} shows the expected rapid increase of $a_1$ with dipole size $r$ at fixed $b$. It levels off at
about $r\simeq 0.7$~fm and then decreases again towards larger $r$ where the dipole grows as large as the proton and
a perturbative calculation loses validity. This behavior is qualitatively similar to the one obtained for the real part of the $S$-matrix in a similar calculation in Ref.~\cite{Dumitru:2021tvw}. These results are not particularly sensitive to the collinear cutoff: using $m_\mathrm{col}=0.3\,\mathrm{GeV}$ instead of $0.2\,\mathrm{GeV}$ results in 5\% (20\%) larger scattering amplitude at small (large) $r$.

It is interesting to compare the typical magnitude of the Odderon
exchange amplitude obtained here to parameterizations commonly employed in the literature as initial conditions at $x\simeq0.01$ for
small-$x$ evolution. Fig.~4 of Ref.~\cite{Lappi:2016gqe}, for example, depicts Odderon amplitudes which reach maximum
values of $\approx 0.15$ and $0.4$, respectively. The initial ``spin dependent Odderon'' amplitude of Refs.~\cite{Yao:2018vcg,Hagiwara:2020mqb}
coincides with the first model of Ref.~\cite{Lappi:2016gqe}.
The maximal (over angle $\theta$ and dipole size $r$) value for the Odderon that we obtain at $x\gsim 0.01$ is about
$5\cdot10^{-3}$ for $\as=0.2$ used in this work. 
On the other hand, the quasi-classical Odderon amplitude derived for a large nucleus, Eq.~(56)
of Ref.~\cite{Kovchegov:2012ga} (also see~\cite{Kovchegov:2003dm,Jeon:2005cf,Zhou:2013gsa}), if applied to a proton (at $r=2b=0.7$~fm) with Gaussian transverse ``profile function''~\cite{Kowalski:2006hc}, is smaller than our result by about one order of magnitude.

\begin{figure}
    \centering
    \includegraphics[width=\columnwidth]{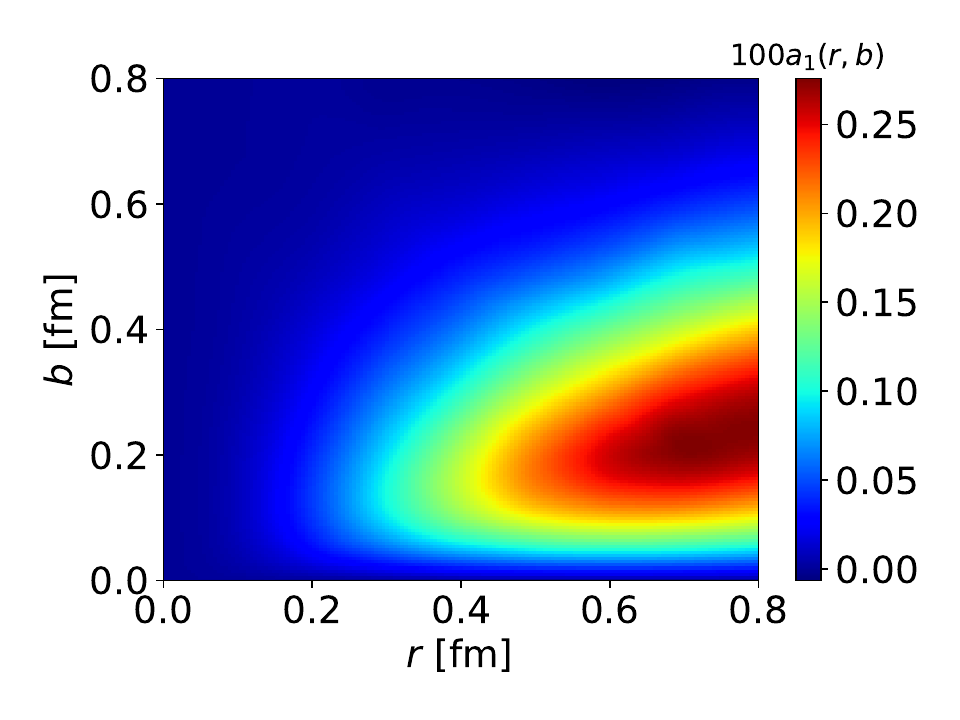}
    \caption{Odderon modulation coefficient $a_1$ as a function of $r$ and $b$ at $x=0.1$ calculated at NLO accuracy.}
    \label{fig:2d_nlo_01}
\end{figure}

\begin{figure}
    \centering
    \includegraphics[width=\columnwidth]{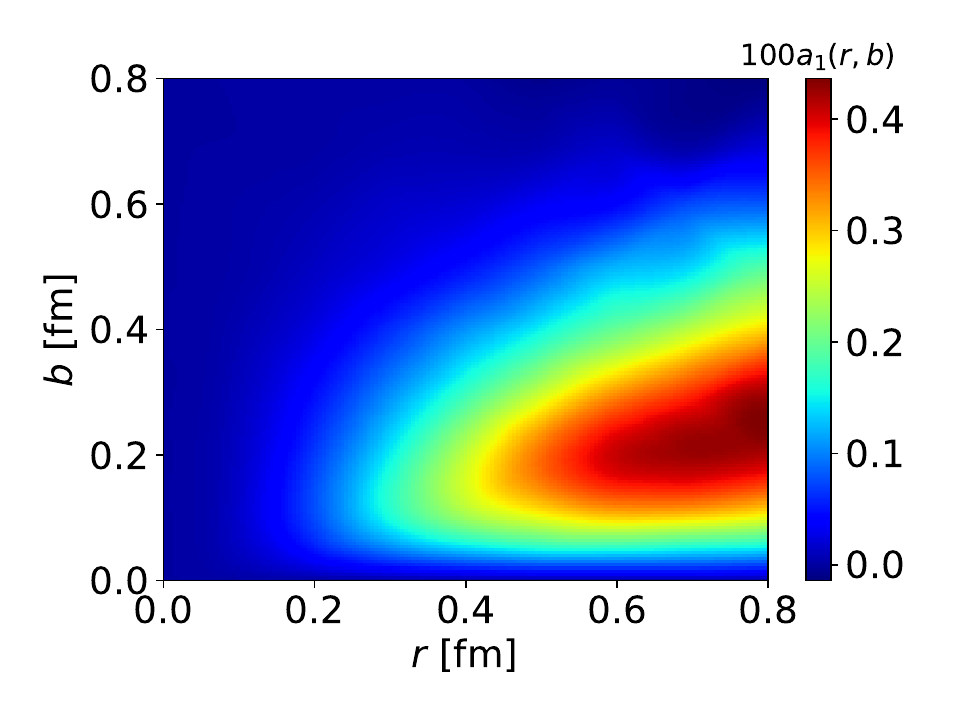}
    \caption{Odderon modulation coefficient $a_1$ as a function of $r$ and $b$ at $x=0.03$ calculated at NLO accuracy. Note that the color scheme is different than in Fig.~\ref{fig:2d_nlo_01}.}
    \label{fig:2d_nlo_003}
\end{figure}

Finally, we illustrate the dominant $a_1$ modulation coefficient at NLO as a function of both $r$ and $b$ in Fig.~\ref{fig:2d_nlo_01} for $x=0.1$ and in Fig.~\ref{fig:2d_nlo_003} for $x=0.03$.  Aside from the increasing magnitude, there is no clear qualitative change in the shape of the Odderon amplitude. At large $b$ the $a_1$ coefficient also changes sign which is visible in these figures. 
In the supplementary material we provide tables for the $a_1$ and $a_3$ coefficients (which are interpolated when generating figures \ref{fig:2d_nlo_01} and \ref{fig:2d_nlo_003}) as functions of $r$ and $b$ at $x=0.1, 0.03$ and $x=0.01$, and for comparison also for the LO three quark proton wave function.

\section{Discussion}

We have presented for the first time an estimate for the perturbative, $C$-odd, dipole-proton three gluon exchange amplitude $O(\vec r, \vec b)$ at moderately small longitudinal momentum fraction $x$ where the target proton includes a perturbative gluon on top of a non-perturbative three-quark Fock state. This is a necessary input for the perturbative small-$x$ evolution of the Odderon. 
We find that $O(\vec r, \vec b)$ increases when the $|qqqg\rangle$ Fock state is added as the number of diagrams increases by an order of magnitude.
Once the proton contains a sufficient number of color charges, the
average dipole $S$-matrix at rapidity $Y=\log x_0/x$ will be given by an
average over the configurations of $A^+$ in the proton:
\be
S_Y(\vec x, \vec y) = \int {\rm D}A^+\, W_Y[A^+]\,
\frac{1}{N_c}\mathrm{tr}\, U(\vec x) U^\dagger(\vec y) ~.
\ee
Here $W_Y[A^+]$ is the weight functional at evolution rapidity $Y$, and $x_0$ is the longitudinal momentum fraction at the initial condition. A small step towards
lower $x$ allows for the emission of an additional soft gluon,
resulting in a small change of $W_Y[A^+]$, i.e.\ the small-$x$ renormalization
group (RG) flow~\cite{Balitsky:1995ub,Balitsky:1998ya,Balitsky:1998kc,Kovchegov:1999yj,Kovchegov:1999ua,Jalilian-Marian:1997jx,Jalilian-Marian:1997gr,JalilianMarian:1997dw,Weigert:2000gi,Iancu:2000hn,Ferreiro:2001qy,Iancu:2001ad,Blaizot:2002xy}.

For weak scattering the average value of $1-S$ is small and the
evolution of the imaginary part $O$ is given
by~\cite{Kovchegov:2003dm,Hatta:2005as,Lappi:2016gqe}
\bea
\partial_Y O(\vec x,\vec y) &=& \frac{\alpha_s N_c}{2\pi^2} \int \mathrm{d}^2 \vec z \,
\frac{(\vec x - \vec y)^2}{(\vec x - \vec z)^2\, (\vec z - \vec y)^2}\nonumber\\
& &
\left[ O(\vec x,\vec z) + O(\vec z,\vec y) - O(\vec x,\vec y)\right] ~.
\eea
For small $r$ the first two terms largely cancel, leaving the
negative virtual correction and a {\em decreasing} Odderon amplitude
with decreasing $x$. (For asymptotically small $x$ the above evolution equation
leads to~\cite{Kovchegov:2003dm} the energy independent
Bartels-Lipatov-Vacca Odderon~\cite{Bartels:1999yt}.)
The observation of such behavior would
indicate the onset of the universal flow predicted by the small-$x$ RG. Our
analysis provides a lower bound on the number of
pre-populated Fock states.

The angular dependence of the Odderon amplitude is found to be well described by $\cos \phi_{\vec r \vec b}$, 
with a small correction proportional to $\cos 3 \phi_{\vec r \vec b}$ which is significant only
in the region where $O(\vec r, \vec b)$ is very small. 
The small magnitude of the perturbative Odderon amplitude obtained here indicates that high luminosities available e.g.\ at the EIC are necessary to access the Odderon experimentally. For example, Ref.~\cite{Dumitru:2019qec}
obtained $\dd\sigma/\dd t \simeq 40$~fb/GeV$^2$ for exclusive $\eta_c$ production in DIS at low $Q^2$, $|t|=1.5$~GeV$^2$, $x=0.1$, in
the LO approximation with $\as=0.35$. 
We intend to compute cross sections for various physical processes from our dipole $S$-matrix in the future.

\subsection*{Acknowledgments}
We thank Y.~Hatta, A.~Kovner and V.~Skokov for useful comments and T. Stebel for noticing a sign error in the original version of this paper. A.D. acknowledges support by the DOE Office of Nuclear Physics through Grant DE-SC0002307, and The City University of New York for PSC-CUNY Research grant 65079-00~53. This work was supported by the Academy of Finland, the Centre of Excellence in Quark Matter and projects 338263 and 346567 (H.M), and projects 347499 and 353772 (R.P). H.M is also supported under the European Union’s Horizon 2020 research and innovation programme by the European Research Council (ERC, grant agreement No. ERC-2018-ADG-835105 YoctoLHC) and by the STRONG-2020 project (grant agreement No. 824093). The content of this article does not reflect the official opinion of the European Union and responsibility for the information and views expressed therein lies entirely with the authors. Computing resources from CSC – IT Center for Science in Espoo, Finland and from the Finnish Grid and Cloud Infrastructure (persistent identifier \texttt{urn:nbn:fi:research-infras-2016072533}) were used in this work.

\appendix

\bibliography{spires}

\providecommand{\href}[2]{#2}\begingroup\raggedright\begin{thebibliography}{10}

\bibitem{Dumitru:2021tqp}
A.~Dumitru, H.~M\"antysaari and R.~Paatelainen, {\it {Cubic color charge
  correlator in a proton made of three quarks and a gluon}},
  \href{http://dx.doi.org/10.1103/PhysRevD.105.036007}{{\em Phys. Rev. D} {\bf
  105} (2022)~no.~3 036007} [\href{http://arXiv.org/abs/2106.12623}{{\tt
  arXiv:2106.12623 [hep-ph]}}].

\bibitem{Lappi:2016gqe}
T.~Lappi, A.~Ramnath, K.~Rummukainen and H.~Weigert, {\it {JIMWLK evolution of
  the odderon}},  \href{http://dx.doi.org/10.1103/PhysRevD.94.054014}{{\em
  Phys. Rev. D} {\bf 94} (2016)~no.~5 054014}
  [\href{http://arXiv.org/abs/1606.00551}{{\tt arXiv:1606.00551 [hep-ph]}}].

\bibitem{Nikolaev:1990ja}
N.~N. Nikolaev and B.~G. Zakharov, {\it Colour transparency and scaling
  properties of nuclear shadowing in deep inelastic scattering},
  \href{http://dx.doi.org/10.1007/BF01483577}{{\em Z. Phys.} {\bf C49} (1991)
  607}.

\bibitem{Mueller:1994gb}
A.~H. Mueller, {\it {Unitarity and the BFKL pomeron}},
  \href{http://dx.doi.org/10.1016/0550-3213(94)00480-3}{{\em Nucl. Phys. B}
  {\bf 437} (1995) 107} [\href{http://arXiv.org/abs/hep-ph/9408245}{{\tt
  arXiv:hep-ph/9408245}}].

\bibitem{Mueller:2001fv}
A.~H. Mueller in {\em Cargese 2001, {QCD} perspectives on hot and dense
  matter}, pp.~45--72.
\newblock 2001.
\newblock \href{http://arXiv.org/abs/hep-ph/0111244}{{\tt
  arXiv:hep-ph/0111244}}.

\bibitem{Kovchegov:2012mbw}
Y.~V. Kovchegov and E.~Levin, {\em {Quantum chromodynamics at high energy}},
  vol.~33.
\newblock Cambridge University Press, 8, 2012.

\bibitem{Brodsky:1997de}
S.~J. Brodsky, H.-C. Pauli and S.~S. Pinsky, {\it Quantum chromodynamics and
  other field theories on the light cone},
  \href{http://dx.doi.org/10.1016/S0370-1573(97)00089-6}{{\em Phys. Rept.} {\bf
  301} (1998) 299} [\href{http://arXiv.org/abs/hep-ph/9705477}{{\tt
  arXiv:hep-ph/9705477 [hep-ph]}}].

\bibitem{Kovchegov:2013cva}
Y.~V. Kovchegov and M.~D. Sievert, {\it {Sivers} function in the quasiclassical
  approximation},  \href{http://dx.doi.org/10.1103/PhysRevD.89.054035}{{\em
  Phys. Rev.} {\bf D89} (2014) 054035}
  [\href{http://arXiv.org/abs/1310.5028}{{\tt arXiv:1310.5028 [hep-ph]}}].

\bibitem{Yao:2018vcg}
X.~Yao, Y.~Hagiwara and Y.~Hatta, {\it {Computing the gluon Sivers function at
  small-$x$}},  \href{http://dx.doi.org/10.1016/j.physletb.2019.01.029}{{\em
  Phys. Lett. B} {\bf 790} (2019) 361}
  [\href{http://arXiv.org/abs/1812.03959}{{\tt arXiv:1812.03959 [hep-ph]}}].

\bibitem{Zhou:2013gsa}
J.~Zhou, {\it {Transverse single spin asymmetries at small x and the anomalous
  magnetic moment}},  \href{http://dx.doi.org/10.1103/PhysRevD.89.074050}{{\em
  Phys. Rev. D} {\bf 89} (2014)~no.~7 074050}
  [\href{http://arXiv.org/abs/1308.5912}{{\tt arXiv:1308.5912 [hep-ph]}}].

\bibitem{Boer:2015pni}
D.~Boer, M.~G. Echevarria, P.~Mulders and J.~Zhou, {\it {Single spin
  asymmetries from a single Wilson loop}},
  \href{http://dx.doi.org/10.1103/PhysRevLett.116.122001}{{\em Phys. Rev.
  Lett.} {\bf 116} (2016)~no.~12 122001}
  [\href{http://arXiv.org/abs/1511.03485}{{\tt arXiv:1511.03485 [hep-ph]}}].

\bibitem{Boussarie:2019vmk}
R.~Boussarie, Y.~Hatta, L.~Szymanowski and S.~Wallon, {\it {Probing the Gluon
  Sivers Function with an Unpolarized Target: GTMD Distributions and the
  Odderons}},  \href{http://dx.doi.org/10.1103/PhysRevLett.124.172501}{{\em
  Phys. Rev. Lett.} {\bf 124} (2020)~no.~17 172501}
  [\href{http://arXiv.org/abs/1912.08182}{{\tt arXiv:1912.08182 [hep-ph]}}].

\bibitem{Hagiwara:2020mqb}
Y.~Hagiwara, Y.~Hatta, R.~Pasechnik and J.~Zhou, {\it {Spin-dependent Pomeron
  and Odderon in elastic proton-proton scattering}},
  \href{http://dx.doi.org/10.1140/epjc/s10052-020-8007-6}{{\em Eur. Phys. J. C}
  {\bf 80} (2020)~no.~5 427} [\href{http://arXiv.org/abs/2003.03680}{{\tt
  arXiv:2003.03680 [hep-ph]}}].

\bibitem{Boer:2022njw}
D.~Boer, Y.~Hagiwara, J.~Zhou and Y.-j. Zhou, {\it {Scale evolution of T-odd
  gluon TMDs at small x}},
  \href{http://dx.doi.org/10.1103/PhysRevD.105.096017}{{\em Phys. Rev. D} {\bf
  105} (2022)~no.~9 096017} [\href{http://arXiv.org/abs/2203.00267}{{\tt
  arXiv:2203.00267 [hep-ph]}}].

\bibitem{Dumitru:2018vpr}
A.~Dumitru, G.~A. Miller and R.~Venugopalan, {\it {Extracting many-body color
  charge correlators in the proton from exclusive DIS at large Bjorken $x$}},
  \href{http://dx.doi.org/10.1103/PhysRevD.98.094004}{{\em Phys. Rev. D} {\bf
  98} (2018)~no.~9 094004} [\href{http://arXiv.org/abs/1808.02501}{{\tt
  arXiv:1808.02501 [hep-ph]}}].

\bibitem{Dumitru:2020gla}
A.~Dumitru and R.~Paatelainen, {\it {Sub-femtometer scale color charge
  fluctuations in a proton made of three quarks and a gluon}},
  \href{http://dx.doi.org/10.1103/PhysRevD.103.034026}{{\em Phys. Rev. D} {\bf
  103} (2021)~no.~3 034026} [\href{http://arXiv.org/abs/2010.11245}{{\tt
  arXiv:2010.11245 [hep-ph]}}].

\bibitem{Dumitru:2020fdh}
A.~Dumitru, V.~Skokov and T.~Stebel, {\it {Subfemtometer scale color charge
  correlations in the proton}},
  \href{http://dx.doi.org/10.1103/PhysRevD.101.054004}{{\em Phys. Rev. D} {\bf
  101} (2020)~no.~5 054004} [\href{http://arXiv.org/abs/2001.04516}{{\tt
  arXiv:2001.04516 [hep-ph]}}].

\bibitem{Schlumpf:1992vq}
F.~Schlumpf, {\it {Relativistic constituent quark model of electroweak
  properties of baryons}},
  \href{http://dx.doi.org/10.1103/PhysRevD.47.4114}{{\em Phys. Rev. D} {\bf 47}
  (1993) 4114} [\href{http://arXiv.org/abs/hep-ph/9212250}{{\tt
  arXiv:hep-ph/9212250}}].
\newblock [Erratum: Phys.Rev.D 49, 6246 (1994)].

\bibitem{Brodsky:1994fz}
S.~J. Brodsky and F.~Schlumpf, {\it {Wave function independent relations
  between the nucleon axial coupling $g_A$ and the nucleon magnetic moments}},
  \href{http://dx.doi.org/10.1016/0370-2693(94)90525-8}{{\em Phys. Lett. B}
  {\bf 329} (1994) 111} [\href{http://arXiv.org/abs/hep-ph/9402214}{{\tt
  arXiv:hep-ph/9402214}}].

\bibitem{Dumitru:2021tvw}
A.~Dumitru, H.~M\"antysaari and R.~Paatelainen, {\it {Color charge correlations
  in the proton at NLO: Beyond geometry based intuition}},
  \href{http://dx.doi.org/10.1016/j.physletb.2021.136560}{{\em Phys. Lett. B}
  {\bf 820} (2021) 136560} [\href{http://arXiv.org/abs/2103.11682}{{\tt
  arXiv:2103.11682 [hep-ph]}}].

\bibitem{Hagler:2002nh}
P.~H\"agler, B.~Pire, L.~Szymanowski and O.~Teryaev, {\it {Hunting the
  QCD-Odderon in hard diffractive electroproduction of two pions}},
  \href{http://dx.doi.org/10.1016/S0370-2693(02)01736-7}{{\em Phys. Lett. B}
  {\bf 535} (2002) 117} [\href{http://arXiv.org/abs/hep-ph/0202231}{{\tt
  arXiv:hep-ph/0202231}}].
\newblock [Erratum: Phys.Lett.B 540, 324--325 (2002)].

\bibitem{Hagler:2002nf}
P.~H\"agler, B.~Pire, L.~Szymanowski and O.~Teryaev, {\it {Pomeron - odderon
  interference effects in electroproduction of two pions}},
  \href{http://dx.doi.org/10.1140/epjc/s2002-01054-9}{{\em Eur. Phys. J. C}
  {\bf 26} (2002) 261} [\href{http://arXiv.org/abs/hep-ph/0207224}{{\tt
  arXiv:hep-ph/0207224}}].

\bibitem{Czyzewski:1996bv}
J.~Czyzewski, J.~Kwiecinski, L.~Motyka and M.~Sadzikowski, {\it {Exclusive
  $\eta_c$ photoproduction and electroproduction at HERA as a possible probe of
  the odderon singularity in QCD}},
  \href{http://dx.doi.org/10.1016/S0370-2693(97)00249-9}{{\em Phys. Lett. B}
  {\bf 398} (1997) 400} [\href{http://arXiv.org/abs/hep-ph/9611225}{{\tt
  arXiv:hep-ph/9611225}}].
\newblock [Erratum: Phys.Lett.B 411, 402 (1997)].

\bibitem{Engel:1997cga}
R.~Engel, D.~Ivanov, R.~Kirschner and L.~Szymanowski, {\it {Diffractive meson
  production from virtual photons with odd charge - parity exchange}},
  \href{http://dx.doi.org/10.1007/s100520050188}{{\em Eur. Phys. J. C} {\bf 4}
  (1998) 93} [\href{http://arXiv.org/abs/hep-ph/9707362}{{\tt
  arXiv:hep-ph/9707362}}].

\bibitem{Kilian:1997ew}
W.~Kilian and O.~Nachtmann, {\it {Single pseudoscalar meson production in
  diffractive e p scattering}},
  \href{http://dx.doi.org/10.1007/s100520050274}{{\em Eur. Phys. J. C} {\bf 5}
  (1998) 317} [\href{http://arXiv.org/abs/hep-ph/9712371}{{\tt
  arXiv:hep-ph/9712371}}].

\bibitem{Rueter:1998gj}
M.~Rueter, H.~G. Dosch and O.~Nachtmann, {\it {Odd CP contributions to
  diffractive processes}},
  \href{http://dx.doi.org/10.1103/PhysRevD.59.014018}{{\em Phys. Rev. D} {\bf
  59} (1999) 014018} [\href{http://arXiv.org/abs/hep-ph/9806342}{{\tt
  arXiv:hep-ph/9806342}}].

\bibitem{Dumitru:2019qec}
A.~Dumitru and T.~Stebel, {\it {Multiquark matrix elements in the proton and
  three gluon exchange for exclusive $\eta_c$ production in photon-proton
  diffractive scattering}},
  \href{http://dx.doi.org/10.1103/PhysRevD.99.094038}{{\em Phys. Rev. D} {\bf
  99} (2019)~no.~9 094038} [\href{http://arXiv.org/abs/1903.07660}{{\tt
  arXiv:1903.07660 [hep-ph]}}].

\bibitem{Mantysaari:2020lhf}
H.~M\"antysaari, K.~Roy, F.~Salazar and B.~Schenke, {\it {Gluon imaging using
  azimuthal correlations in diffractive scattering at the Electron-Ion
  Collider}},  \href{http://dx.doi.org/10.1103/PhysRevD.103.094026}{{\em Phys.
  Rev. D} {\bf 103} (2021)~no.~9 094026}
  [\href{http://arXiv.org/abs/2011.02464}{{\tt arXiv:2011.02464 [hep-ph]}}].

\bibitem{Bzdak:2007cz}
A.~Bzdak, L.~Motyka, L.~Szymanowski and J.~R. Cudell, {\it {Exclusive
  $\mathrm{J}/\psi$ and Upsilon hadroproduction and the QCD odderon}},
  \href{http://dx.doi.org/10.1103/PhysRevD.75.094023}{{\em Phys. Rev. D} {\bf
  75} (2007) 094023} [\href{http://arXiv.org/abs/hep-ph/0702134}{{\tt
  arXiv:hep-ph/0702134}}].

\bibitem{Kovchegov:2003dm}
Y.~V. Kovchegov, L.~Szymanowski and S.~Wallon, {\it {Perturbative odderon in
  the dipole model}},
  \href{http://dx.doi.org/10.1016/j.physletb.2004.02.036}{{\em Phys. Lett. B}
  {\bf 586} (2004) 267} [\href{http://arXiv.org/abs/hep-ph/0309281}{{\tt
  arXiv:hep-ph/0309281}}].

\bibitem{Hatta:2005as}
Y.~Hatta, E.~Iancu, K.~Itakura and L.~McLerran, {\it {Odderon in the color
  glass condensate}},
  \href{http://dx.doi.org/10.1016/j.nuclphysa.2005.05.163}{{\em Nucl. Phys. A}
  {\bf 760} (2005) 172} [\href{http://arXiv.org/abs/hep-ph/0501171}{{\tt
  arXiv:hep-ph/0501171}}].

\bibitem{Accardi:2012qut}
A.~Accardi {\em et.~al.}, {\it {Electron Ion Collider: The Next QCD Frontier}:
  {Understanding the glue that binds us all}},
  \href{http://dx.doi.org/10.1140/epja/i2016-16268-9}{{\em Eur. Phys. J. A}
  {\bf 52} (2016)~no.~9 268} [\href{http://arXiv.org/abs/1212.1701}{{\tt
  arXiv:1212.1701 [nucl-ex]}}].

\bibitem{Aschenauer:2017jsk}
E.~Aschenauer, S.~Fazio, J.~Lee, H.~Mäntysaari, B.~Page, B.~Schenke,
  T.~Ullrich, R.~Venugopalan and P.~Zurita, {\it {The electron\textendash{}ion
  collider: assessing the energy dependence of key measurements}},
  \href{http://dx.doi.org/10.1088/1361-6633/aaf216}{{\em Rept. Prog. Phys.}
  {\bf 82} (2019)~no.~2 024301} [\href{http://arXiv.org/abs/1708.01527}{{\tt
  arXiv:1708.01527 [nucl-ex]}}].

\bibitem{AbdulKhalek:2021gbh}
R.~Abdul~Khalek {\em et.~al.}, {\it {Science Requirements and Detector Concepts
  for the Electron-Ion Collider}: {EIC Yellow Report}},
  \href{http://dx.doi.org/10.1016/j.nuclphysa.2022.122447}{{\em Nucl. Phys. A}
  {\bf 1026} (2022) 122447} [\href{http://arXiv.org/abs/2103.05419}{{\tt
  arXiv:2103.05419 [physics.ins-det]}}].

\bibitem{Kowalski:2006hc}
H.~Kowalski, L.~Motyka and G.~Watt, {\it Exclusive diffractive processes at
  {HERA} within the dipole picture},
  \href{http://dx.doi.org/10.1103/PhysRevD.74.074016}{{\em Phys. Rev.} {\bf
  D74} (2006) 074016} [\href{http://arXiv.org/abs/hep-ph/0606272}{{\tt
  arXiv:hep-ph/0606272}}].

\bibitem{Kovchegov:2012ga}
Y.~V. Kovchegov and M.~D. Sievert, {\it {A New Mechanism for Generating a
  Single Transverse Spin Asymmetry}},
  \href{http://dx.doi.org/10.1103/PhysRevD.86.034028}{{\em Phys. Rev. D} {\bf
  86} (2012) 034028} [\href{http://arXiv.org/abs/1201.5890}{{\tt
  arXiv:1201.5890 [hep-ph]}}].
\newblock [Erratum: Phys.Rev.D 86, 079906 (2012)].

\bibitem{Jeon:2005cf}
S.~Jeon and R.~Venugopalan, {\it A classical odderon in {QCD} at high
  energies},  \href{http://dx.doi.org/10.1103/PhysRevD.71.125003}{{\em Phys.
  Rev.} {\bf D71} (2005) 125003}
  [\href{http://arXiv.org/abs/hep-ph/0503219}{{\tt arXiv:hep-ph/0503219
  [hep-ph]}}].

\bibitem{Balitsky:1995ub}
I.~Balitsky, {\it Operator expansion for high-energy scattering},
  \href{http://dx.doi.org/10.1016/0550-3213(95)00638-9}{{\em Nucl. Phys.} {\bf
  B463} (1996) 99} [\href{http://arXiv.org/abs/hep-ph/9509348}{{\tt
  arXiv:hep-ph/9509348}}].

\bibitem{Balitsky:1998ya}
I.~Balitsky, {\it Factorization and high-energy effective action},
  \href{http://dx.doi.org/10.1103/PhysRevD.60.014020}{{\em Phys. Rev.} {\bf
  D60} (1999) 014020} [\href{http://arXiv.org/abs/hep-ph/9812311}{{\tt
  arXiv:hep-ph/9812311}}].

\bibitem{Balitsky:1998kc}
I.~Balitsky, {\it Factorization for high-energy scattering},
  \href{http://dx.doi.org/10.1103/PhysRevLett.81.2024}{{\em Phys. Rev. Lett.}
  {\bf 81} (1998) 2024} [\href{http://arXiv.org/abs/hep-ph/9807434}{{\tt
  arXiv:hep-ph/9807434}}].

\bibitem{Kovchegov:1999yj}
Y.~V. Kovchegov, {\it {Small $x$ $F_2$ structure function of a nucleus
  including multiple pomeron exchanges}},
  \href{http://dx.doi.org/10.1103/PhysRevD.60.034008}{{\em Phys. Rev. D} {\bf
  60} (1999) 034008} [\href{http://arXiv.org/abs/hep-ph/9901281}{{\tt
  arXiv:hep-ph/9901281}}].

\bibitem{Kovchegov:1999ua}
Y.~V. Kovchegov, {\it Unitarization of the {BFKL} pomeron on a nucleus},
  \href{http://dx.doi.org/10.1103/PhysRevD.61.074018}{{\em Phys. Rev.} {\bf
  D61} (2000) 074018} [\href{http://arXiv.org/abs/hep-ph/9905214}{{\tt
  arXiv:hep-ph/9905214}}].

\bibitem{Jalilian-Marian:1997jx}
J.~Jalilian-Marian, A.~Kovner, A.~Leonidov and H.~Weigert, {\it The {BFKL}
  equation from the {Wilson} renormalization group},
  \href{http://dx.doi.org/10.1016/S0550-3213(97)00440-9}{{\em Nucl. Phys.} {\bf
  B504} (1997) 415} [\href{http://arXiv.org/abs/hep-ph/9701284}{{\tt
  arXiv:hep-ph/9701284}}].

\bibitem{Jalilian-Marian:1997gr}
J.~Jalilian-Marian, A.~Kovner, A.~Leonidov and H.~Weigert, {\it The {Wilson}
  renormalization group for low x physics: Towards the high density regime},
  \href{http://dx.doi.org/10.1103/PhysRevD.59.014014}{{\em Phys. Rev.} {\bf
  D59} (1999) 014014} [\href{http://arXiv.org/abs/hep-ph/9706377}{{\tt
  arXiv:hep-ph/9706377}}].

\bibitem{JalilianMarian:1997dw}
J.~Jalilian-Marian, A.~Kovner and H.~Weigert, {\it The {Wilson} renormalization
  group for low x physics: Gluon evolution at finite parton density},
  \href{http://dx.doi.org/10.1103/PhysRevD.59.014015}{{\em Phys. Rev.} {\bf
  D59} (1999) 014015} [\href{http://arXiv.org/abs/hep-ph/9709432}{{\tt
  arXiv:hep-ph/9709432}}].

\bibitem{Weigert:2000gi}
H.~Weigert, {\it Unitarity at small {Bjorken} x},
  \href{http://dx.doi.org/10.1016/S0375-9474(01)01668-2}{{\em Nucl. Phys.} {\bf
  A703} (2002) 823} [\href{http://arXiv.org/abs/hep-ph/0004044}{{\tt
  arXiv:hep-ph/0004044 [hep-ph]}}].

\bibitem{Iancu:2000hn}
E.~Iancu, A.~Leonidov and L.~D. McLerran, {\it Nonlinear gluon evolution in the
  color glass condensate. i},
  \href{http://dx.doi.org/10.1016/S0375-9474(01)00642-X}{{\em Nucl. Phys.} {\bf
  A692} (2001) 583} [\href{http://arXiv.org/abs/hep-ph/0011241}{{\tt
  arXiv:hep-ph/0011241}}].

\bibitem{Ferreiro:2001qy}
E.~Ferreiro, E.~Iancu, A.~Leonidov and L.~McLerran, {\it Nonlinear gluon
  evolution in the color glass condensate. ii},
  \href{http://dx.doi.org/10.1016/S0375-9474(01)01329-X}{{\em Nucl. Phys.} {\bf
  A703} (2002) 489} [\href{http://arXiv.org/abs/hep-ph/0109115}{{\tt
  arXiv:hep-ph/0109115}}].

\bibitem{Iancu:2001ad}
E.~Iancu, A.~Leonidov and L.~D. McLerran, {\it The renormalization group
  equation for the color glass condensate},
  \href{http://dx.doi.org/10.1016/S0370-2693(01)00524-X}{{\em Phys. Lett.} {\bf
  B510} (2001) 133} [\href{http://arXiv.org/abs/hep-ph/0102009}{{\tt
  arXiv:hep-ph/0102009}}].

\bibitem{Blaizot:2002xy}
J.-P. Blaizot, E.~Iancu and H.~Weigert, {\it Nonlinear gluon evolution in path
  integral form},  \href{http://dx.doi.org/10.1016/S0375-9474(02)01299-X}{{\em
  Nucl. Phys.} {\bf A713} (2003) 441}
  [\href{http://arXiv.org/abs/hep-ph/0206279}{{\tt arXiv:hep-ph/0206279
  [hep-ph]}}].

\bibitem{Bartels:1999yt}
J.~Bartels, L.~N. Lipatov and G.~P. Vacca, {\it {A New odderon solution in
  perturbative QCD}},
  \href{http://dx.doi.org/10.1016/S0370-2693(00)00221-5}{{\em Phys. Lett. B}
  {\bf 477} (2000) 178} [\href{http://arXiv.org/abs/hep-ph/9912423}{{\tt
  arXiv:hep-ph/9912423}}].

\end{thebibliography}\endgroup
\bibliographystyle{JHEP-2modlong}

\end{document}